\pdfoutput=1
\documentclass[draft]{agujournal}
\draftfalse

\usepackage{epstopdf}
\usepackage{url}

\journalname{JGR-Earth Surface}

\begin{document}
\title{Large Effects of Particle Size Heterogeneity on Dynamic Saltation Threshold}
\authors{Wei Zhu\affil{1}, Xinghui Huo\affil{2}, Jie Zhang\affil{2}, Peng Wang\affil{2}, Thomas P\"ahtz\affil{1,3}, Ning Huang\affil{2}, and Zhiguo He\affil{1,3}}

\affiliation{1}{Institute of Port, Coastal and Offshore Engineering, Ocean College, Zhejiang University, 866 Yu Hang Tang Road, 310058 Hangzhou, China}
\affiliation{2}{Key Laboratory of Mechanics on Disaster and Environment in Western China, the Ministry of Education of China, College of Civil Engineering and Mechanics, Lanzhou University, Lanzhou 730000, China}
\affiliation{3}{State Key Laboratory of Satellite Ocean Environment Dynamics, Second Institute of Oceanography, 36 North Baochu Road, 310012 Hangzhou, China}

\correspondingauthor{Thomas P\"ahtz}{0012136@zju.edu.cn}
\correspondingauthor{Ning Huang}{huangn@lzu.edu.cn}

\begin{keypoints}
\item The dynamic threshold of aeolian transport can be much larger for heterogeneous than for homogeneous sand beds with the same median diameter
\item For sufficiently heterogeneous sand beds, there is more than one dynamic aeolian transport threshold because of particle size effects
\item For sufficiently heterogeneous sand beds near the smallest dynamic threshold, aerodynamic entrainment seems to sustain saltation transport
\end{keypoints}

\begin{abstract}
Reliably predicting the geomorphology and climate of planetary bodies requires knowledge of the dynamic threshold wind shear velocity below which saltation transport ceases. Here we measure this threshold in a wind tunnel for four well-sorted and two poorly sorted sand beds by visual means and by a method that exploits a regime shift in the behavior of the surface roughness caused by momentum transfer from the wind to the saltating particles. For our poorly sorted sands, we find that these measurement methods yield different threshold values because, at the smaller visual threshold, relatively coarse particles do not participate in saltation. We further find that both methods yield threshold values that are much larger (60--250\%) for our poorly sorted sands than for our well-sorted sands with similar median particle diameter. In particular, even a rescaling of the dynamic saltation threshold based on the 90th percentile particle diameter rather than the median diameter cannot fully capture this difference, suggesting that relatively very coarse particles have a considerable control on the dynamic threshold. Similar findings were previously reported for water-driven sediment transport. Our findings have important implications for quantitative predictions of saltation transport-related geophysical processes, such as dust aerosol emission.
\end{abstract}

\section{Introduction}
Saltation, which refers to a ballistic hopping motion of granular particles, is the predominant mode in which sand particles are transported by wind on Earth and other planets and one of the most important processes responsible for the shaping of arid planetary surfaces~\citep{Bourkeetal10,Duranetal11,Koketal12,Merrison12,Rasmussenetal15,Valanceetal15}. On Earth, saltation transport is also the main driver of dust aerosol emission and thus has an important impact on its climate~\citep{Koketal14a,Koketal14b,Koketal18,Hausteinetal15}. A key quantity characterizing saltation transport and its impact on Earth's climate is the dynamic saltation threshold: the minimal value $u_t$ of the wind shear velocity $u_\ast\equiv\sqrt{\tau/\rho_{\mathrm{a}}}$ at which saltation transport can be sustained once initiated, where $\rho_{\mathrm{a}}$ is the air density and $\tau$ the shear stress exerted on the sand bed surface.

In contrast to the static threshold above which saltation can be initiated~\citep[e.g.,][and references therein]{Bagnold37,Chepil45,Gilletteetal80,Iversenetal87,Nickling88,IversenRasmussen94,Merrisonetal07,deVetetal14,Burretal15,Raffaeleetal16}, $u_t$ has only rarely been systematically studied in controlled laboratory settings, especially in recent history (there are numerous poorly controlled field studies though~\citep[][and references therein]{BarchynHugenholtz11,Martinetal13,Lietal14,MartinKok17,MartinKok18}). Even today, we largely rely on the old data sets by \citet{Bagnold37} and \citet{Chepil45}, who measured $u_t$ by visual means. However, while \citet{Bagnold37} and \citet{Chepil45} reported the mean particle diameters of the particles composing their tested sand beds, they did not report the particle size distributions. As a matter of fact, controlled studies on the effect of the particle size distribution on $u_t$ have not been carried out yet. More recently, dynamic thresholds were only reported sporadically as by-products of laboratory studies with focus on different matters~\citep{IversenRasmussen94,Creysselsetal09,Hoetal11,LiMcKennaNeumann12,Carneiroetal15}. For example, \citet{Creysselsetal09} and \citet{Hoetal11} indirectly obtained $u_t$ from extrapolating measurements of the saltation transport rate $Q$ to vanishing $Q$, which is also a standard method applied to field data sets~\citep[][and references therein]{BarchynHugenholtz11,Lietal14,MartinKok17}.

Here we apply the extrapolation method to the data sets by \citet{Creysselsetal09} and \citet{Hoetal11} using two recent transport laws and show that the resulting values of $u_t$ can vary from each other by a factor of up to $1.7$ depending on the fitting procedure (section~\ref{ExtrapolationMethod}). Together with the lacking understanding of the effect of the particle size distribution on $u_t$, this stark discrepancy motivated us to carry out controlled laboratory measurements of $u_t$ (section~\ref{Experiments}).

\section{Extrapolation Method} \label{ExtrapolationMethod}
Based on their experiments and a physical parametrization of near-surface particle dynamics, \citet{Creysselsetal09} proposed a linear relationship between the nondimensionalized transport rate $Q_\ast\equiv Q/\left(\rho_{\mathrm{p}}\sqrt{(\rho_{\mathrm{p}}/\rho_{\mathrm{a}}-1)gd_{50}^3}\right)$ and the \textit{Shields number} $\Theta\equiv\rho_{\mathrm{a}}u_\ast^2/[(\rho_{\mathrm{p}}-\rho_{\mathrm{a}})gd_{50}]$ (the reason for parametrizing aeolian transport by $\Theta$ becomes apparent shortly), where $\rho_{\mathrm{p}}$ is the particle density, $g$ the gravitational constant, and $d_{50}$ the median particle diameter:
\begin{linenomath*}
\begin{equation}
 Q_\ast(\Theta)=C_Q\sqrt{\rho_{\mathrm{a}}/\rho_{\mathrm{p}}}(\Theta-\Theta_t), \label{QCreyssels}
\end{equation}
\end{linenomath*}  
where $\Theta_t=\rho_{\mathrm{a}}u_t^2/[(\rho_{\mathrm{p}}-\rho_{\mathrm{a}})gd_{50}]$ is the dynamic threshold Shields number and $C_Q$ a proportionality factor. Such a linear transport law is currently favored among most aeolian transport physicists~\citep{Creysselsetal09,Hoetal11,Duranetal11,Koketal12,MartinKok17,MartinKok18}. However, discrete element-based simulations of saltation transport suggest a nonlinear transport law because of midair collisions~\citep{Carneiroetal13}, which can be parametrized via (see Figure~S1 in the supporting information, which shows data from numerical discrete element method-based simulations of sediment transport that have been experimentally validated in a number of recent studies~\citep{Duranetal12,Duranetal14a,Duranetal14b,PahtzDuran17,PahtzDuran18a,PahtzDuran18b}):  
\begin{linenomath*}
\begin{equation}
 Q_\ast(\Theta)=2\kappa^{-1}\sqrt{\Theta_t}M(\Theta)\max\left(1,\sqrt{M(\Theta)/M_{\mathrm{c}}}\right)\quad\text{with}\quad M(\Theta)=\mu_{\mathrm{b}}^{-1}(\Theta-\Theta_t), \label{QPahtz}
\end{equation}
\end{linenomath*}
where $\kappa=0.4$ is the von K\'arm\'an constant, $\mu_{\mathrm{b}}=0.63$ the bed friction coefficient (i.e., the ratio between granular shear stress and normal-bed pressure at the sand bed surface), and $M_{\mathrm{c}}=0.13$ the critical value of the nondimensionalized transport load $M(\Theta)$ above which midair collisions become significant in dissipating energy. In fact, equation~(\ref{QPahtz}) is linear in $\Theta$ only for $M\leq M_{\mathrm{c}}$, and the linearity of $M$ with the Shields number $\Theta$ (shown and discussed by \citet{PahtzDuran18b}) is the reason why we have used $\Theta$ to parametrize aeolian transport in the first place.

\begin{figure}[htb]
 \begin{center}
  \includegraphics[width=1.0\columnwidth]{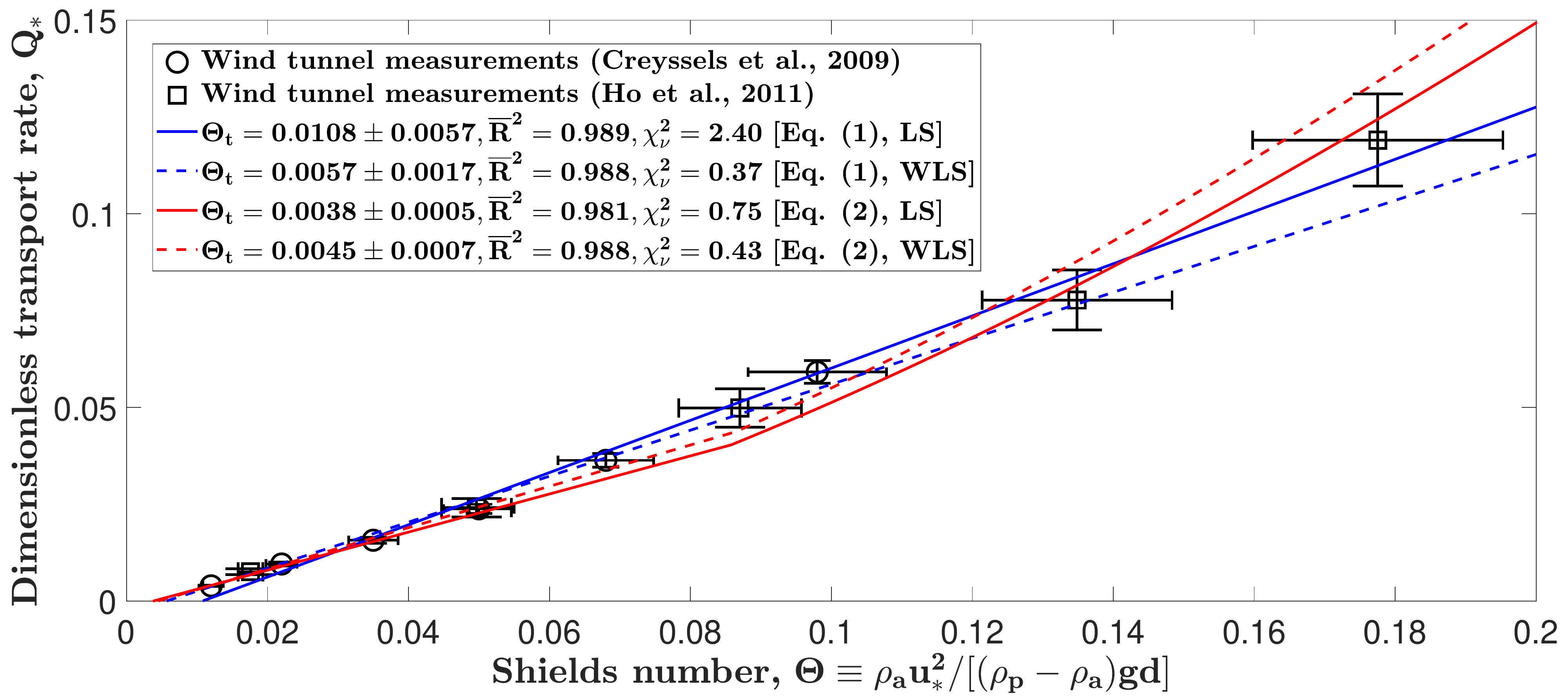}
 \end{center}
 \caption{Comparison of four different extrapolation procedures to determine the dynamic threshold Shields number $\Theta_t$. Nondimensionalized transport rate $Q_\ast\equiv Q/\left(\rho_{\mathrm{p}}\sqrt{(\rho_{\mathrm{p}}/\rho_{\mathrm{a}}-1)gd_{50}^3}\right)$ versus Shields number $\Theta\equiv\rho_{\mathrm{a}}u_\ast^2/[(\rho_{\mathrm{p}}-\rho_{\mathrm{a}})gd_{50}]$. Symbols and error bars correspond to the wind tunnel measurements by \citet[][$\rho_{\mathrm{p}}=2500~$kg/m$^3$, $d_{50}=242~\mu$m]{Creysselsetal09} and \citet[][$\rho_{\mathrm{p}}=2470~$kg/m$^3$, $d_{50}=232~\mu$m]{Hoetal11}. Note that the data by \citet[][figure~2]{Hoetal11} were slightly modified by the leading researcher later on~\citep[][figure~7.4, which is the data shown here]{Ho12}. The uncertainty in the fitted values of $\Theta_t$ and $u_t$ indicates the 95\% confidence interval, which we estimated from assuming that the standard error of these values is the standard deviation of a Student's $t$ distribution with $m-p$ degrees of freedom, where $m$ is the number of measurements and $p$ the number of fit parameters. The adjusted coefficient of determination ($\overline{R}^2$) is calculated from $R^2=1-\sum_iw_i[Q_\ast(\Theta_i)-Q_{\ast i}]^2/\sum_iw_i[\overline{Q}_\ast-Q_{\ast i}]^2$ ($w_i=1$ for unweighted cases) through $\overline{R}^2=1-(1-R^2)(m-1)/(m-p)$, where $\overline{Q}_\ast$ is the weighted mean of $Q_{\ast i}$. The reduced chi-squared is calculated through $\chi_\nu^2=\sum_iw_i[Q_\ast(\Theta_i)-Q_{\ast i}]^2/(m-p)$ (using the same $w_i$ for unweighted cases as for the weighted cases).}
\label{ExtrapolationProcedure}
\end{figure}

We now fit both transport laws to paired wind tunnel measurements of $\Theta$ and $Q_\ast$ by \citet{Creysselsetal09} and \citet{Hoetal11} using $C_Q$ and $\Theta_t$ (equation~(\ref{QCreyssels})) or only $\Theta_t$ (equation~(\ref{QPahtz})) as fit parameters. To carry out the fit, we employ two different fitting procedures: least-squares (i.e., minimization of $\sum_i[Q_\ast(\Theta_i)-Q_{\ast i}]^2$) and weighted least-squares (i.e., minimization of $\sum_iw_i[Q_\ast(\Theta_i)-Q_{\ast i}]^2$, where the weights account for absolute measurement uncertainties: $w_i=[\Delta Q_{\ast i}^2+[Q_\ast^\prime(\Theta_i)\Delta\Theta_i]^2]^{-1}$, the same as method has been used by \citet{MartinKok17}). In contrast to the weighted least-squares procedure, the least-squares procedure effectively assumes $w_i=\mathrm{const}$ and thus constant absolute measurement uncertainties. Because the measurements by \citet{Creysselsetal09} and \citet{Hoetal11} exhibited a constant relative uncertainty $\Delta\Theta_i/\Theta_i$ of $10\%$ and constant relative uncertainties $\Delta Q_{\ast i}/Q_{\ast i}$ of $5\%$ and $10\%$~\citep{Ho12}, respectively, assuming constant absolute measurement uncertainties underweighs near-threshold measurements and overweighs measurements far from the threshold. Given that the goal of the fitting procedure is the estimation of $\Theta_t$, this underweighing of near-threshold measurements can be very problematic. In fact, Figure~\ref{ExtrapolationProcedure} shows that the value of $\Theta_t$ estimated from the data sets by \citet{Creysselsetal09} and \citet{Hoetal11} varies with the applied transport law and fitting procedure by up to a factor of $2.8$, which corresponds to a variability of $u_t$ by a factor of $1.7$, and most of this variability is caused by the least-squares fitting procedure.
We thus conclude that threshold values obtained using the extrapolation method are very unreliable when employing the least-squares fitting procedure (as done by most previous studies) and that the weighted least-squares method should be used instead (as done by \citet{MartinKok17}).

\section{Wind Tunnel Experiments} \label{Experiments}
\subsection{Instrumentation and Experimental Protocol}
\begin{figure}[htb]
 \begin{center}
  \includegraphics[width=1.0\columnwidth]{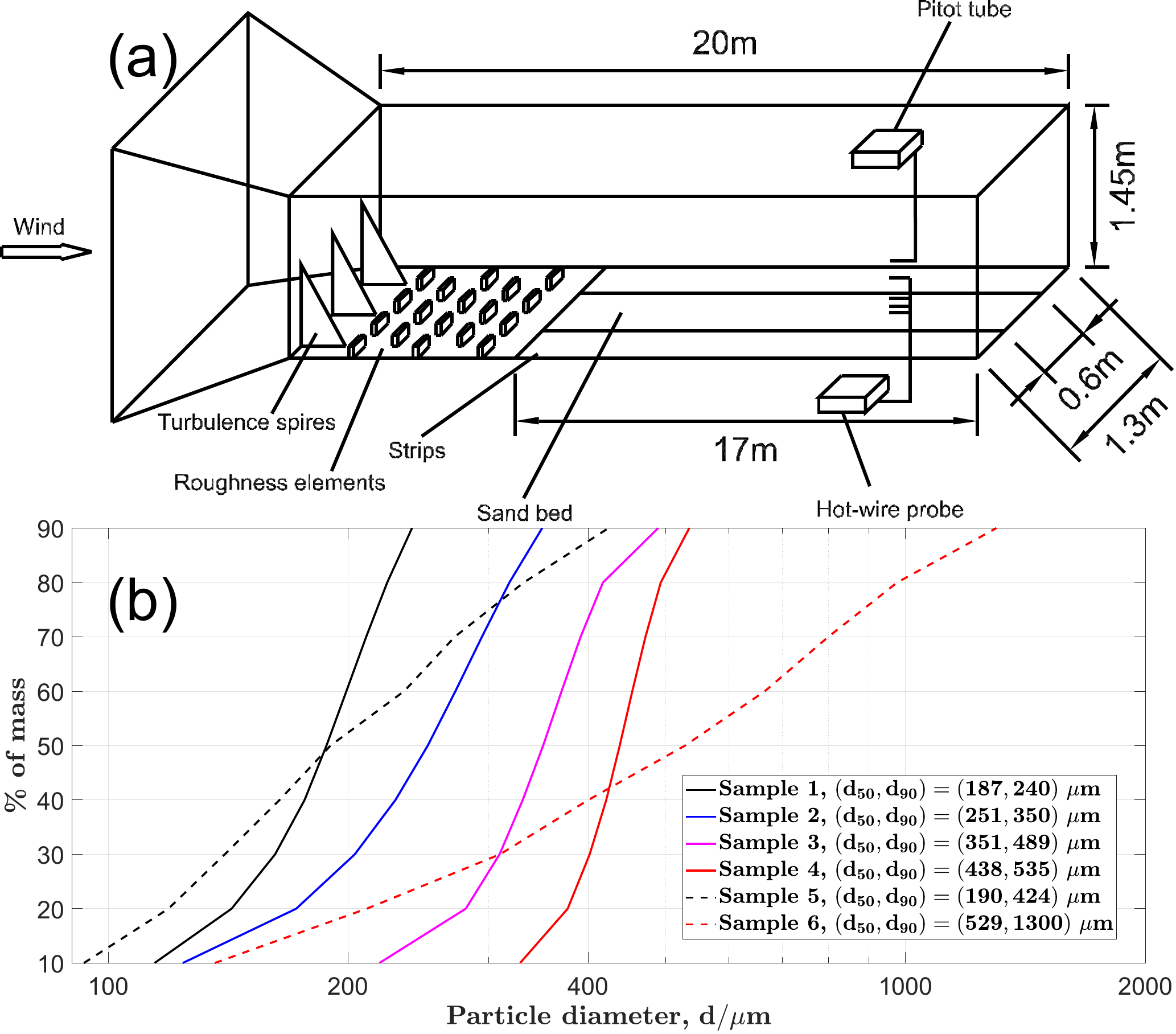}
 \end{center}
 \caption{(a) Sketch of the wind tunnel at Lanzhou University. (b) Particle size distributions of sand beds used in the experiments, where Sample~6 refers to the original sand from Tengger Desert.}
\label{WindTunnel}
\end{figure}

\begin{figure}[htb]
 \begin{center}
  \includegraphics[width=1.0\columnwidth]{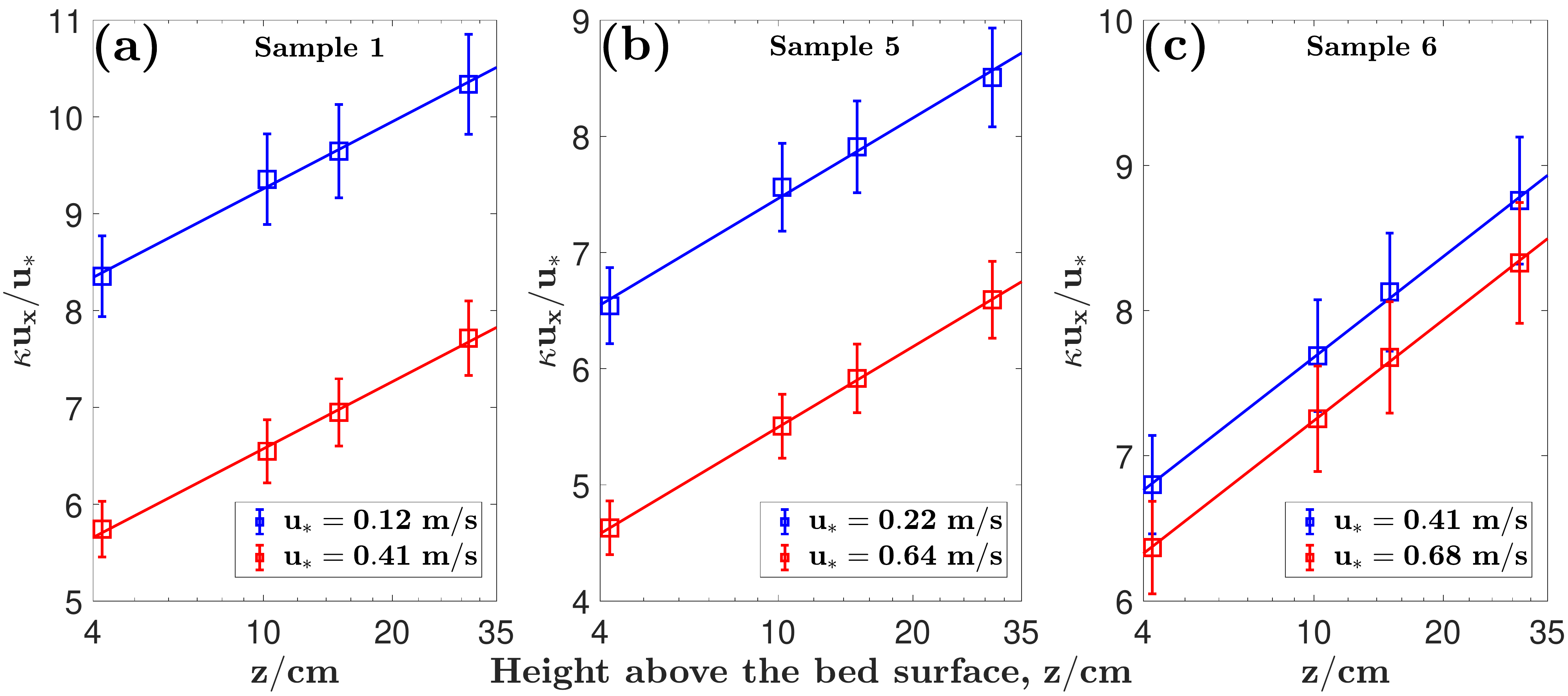}
 \end{center}
 \caption{\added{Exemplary mean wind velocity profiles for (a) sand Sample~1, (b) sand Sample~5, and (c) sand Sample~6. Squares and error bars correspond to the measurements by the hot-wire probes. Lines correspond to equation~(\ref{ulog}) for conditions with absent (blue) and present (red) saltation transport.}}
\label{VelocityProfiles}
\end{figure}

The experiments were conducted in a wind tunnel located at Lanzhou University (Figure~\ref{WindTunnel}a) using an experimental setup similar to that of \citet{Zhangetal14}.
The working section of the wind tunnel was $20$~m long, with a cross-section of $1.3$~m width and $1.45$~m height. Roughness elements and turbulence spires were placed in front of the working section in order to generate a boundary layer that is similar to the one at the downstream end (i.e., the boundary layer development starts with a condition that is not too far from the fully developed state). We used six different sand beds, each about $17$~m long, $0.6$~m wide, and $6$~cm thick: one sand bed consisting of the original sand from Tengger Desert ($\rho_{\mathrm{p}}=2650$~kg/m$^3$) and five differently sieved sand samples (see particle size distribution in Figure~\ref{WindTunnel}b, where ``Sample~6'' refers to the original sand). Each sand sample was flattened to the height of adjacent hard strips before each experimental run. For each sand sample, the free stream wind velocity $U_\infty$, measured by a pitot tube, was successively decremented from a large value corresponding to intense transport to a low value well below the dynamic saltation threshold. The intense conditions at the beginning of each run led to a very rapid formation of downstream migrating ripples, the shape of which became roughly steady within less than $20~$s (detected via illumination with a bright spotlight). For the less intense conditions, the shape of the ripples, which continued to migrate downstream, did not change notably. Sand was not fed at the tunnel entrance because the working section was sufficiently long to ensure saturated transport at its downstream end~\citep{Selmanietal18}. The wind velocity was measured near the end of the working section at four elevations $z$ above the sand bed ($z=[4.2,10.2,15,30.1]$~cm) using I-type hot-wire probes (DANTEC~55P11, accuracy $\pm5\%$), which were connected to constant-temperature hot-wire anemometers. The wind velocity was averaged over a 3-min period (sufficient to capture the entire turbulence frequency spectrum) for each $U_\infty$, which means that the typical duration of a run was about 1~hr (3~min times the number of measured $U_\infty$ per run). The wind shear velocity $u_\ast$ was obtained from fitting the log-law to the averaged data ($u_x$):
\begin{linenomath*}
\begin{equation}
 u_x(z)=\frac{u_\ast}{\kappa}\ln\frac{z}{z_o}, \label{ulog}
\end{equation}
\end{linenomath*}
where $z_o$ is the roughness of the sand bed. To confirm that transport was saturated, we also carried out a few test measurements further upwind: The velocity profiles were nearly the same. Although one may expect influences from the side walls and a slight velocity wake at the largest elevation ($z=30.1~$cm), the measurements usually obeyed a logarithmic behavior within the error bars (see Figure~\ref{VelocityProfiles} for exemplary mean wind velocity profiles).
However, we confirmed for a few test cases that deviations from equation~(\ref{ulog}) occur for elevations with $z>30.1~$cm.

\subsection{Threshold Measurement Methods}
\subsubsection{Visual Method} \label{VisualMethod}
The visual method is a standard method to determine the dynamic threshold~\citep{Bagnold37,Chepil45,IversenRasmussen94,LiMcKennaNeumann12,Carneiroetal15}. A bright spotlight illuminated the wind tunnel at the end of the working section, which allowed us to judge whether or not saltation transport was occurring from a three-dimensional perspective. In fact, from this perspective, there was a very clear change in the transport activity from widespread transport to (nearly) no transport in all our experimental runs when decrementing $U_\infty$ from a certain value $U^\mathrm{trans}_\infty$ to the next lower value $U^\mathrm{notrans}_\infty$. We believe that this clear-cut change is essentially the same as the ``sand cloud effect'' described by \citet[][pp. 32--33]{Bagnold41}. We defined the associated visual threshold as the arithmetic mean of the wind shear velocities corresponding to these two free stream velocities: $u^{\mathrm{vis}}_t\equiv(u^\mathrm{trans}_\ast+u^\mathrm{notrans}_\ast)/2$. This definition may lead to a slight systematic uncertainty (as we did not determine the exact shear velocity at which transport stopped), which plays a role for the interpretation of our measurements in section~\ref{VisualvsRoughness}. Note that we also carried out an experimental test run using a static threshold protocol (i.e., successively incrementing instead of decrementing the wind speed) for Sample~1. The visual threshold obtained from this run was considerably larger than those obtained from our standard runs for Sample~1, which confirms that our visual method truly determines a \textit{dynamic} threshold despite the absence of sand feeding in our experiments.

\subsubsection{Roughness Method} \label{RoughnessMethod}
Saltation transport laws, such as equations~(\ref{QCreyssels}) and (\ref{QPahtz}), assume that saltation transport is saturated (or continuous)~\citep{PahtzDuran18a}. That is, if we defined $u_t$ indirectly through a saltation transport law (which is the assumed definition whenever one uses a saltation transport law to predict $Q$), $u_t$ should convey information about the saturated state even though transport near $u_t$ is intermittent and thus undersaturated~\citep{MartinKok18}. According to recent studies~\citep{PahtzDuran18a,PahtzDuran18b}, transport saturates because splash entrainment of bed sediment supplies the transport layer nearly continuously with bed sediment until the flow becomes so strongly suppressed by the negative feedback of the particle motion that it can no longer compensate energy losses of rebounding particles, resulting in a sudden strong increase of deposition that compensates splash entrainment. Consistent with this hypothesis, experiments revealed that, for saturated transport, the local wind velocities near the surface decrease with $u_\ast$ because of this feedback~\citep{Walteretal14}. Experiments further revealed that, above the region of strongly suppressed near-surface wind, there is a focal region at which the wind velocities are nearly constant with $u_\ast$ (the \textit{Bagnold focus}): $u_f\simeq\kappa^{-1}u_\ast\ln(z_f/z_o)$~\citep{Bagnold36}, where $z_f$ and $u_f$ are constants. This focal point approximation is equivalent to an exponentially increasing surface roughness, $z_o=z_f\exp(-\kappa u_f/u_\ast)$, as found in simulations~\citep{Duranetal11,Duranetal12} and measurements~\citep{Creysselsetal09,Hoetal11,Martinetal13} of saturated saltation transport. In contrast, $z_o$ changes much more slowly with $u_\ast$ when saltation transport is strongly undersaturated or absent: $z_o\simeq\mathrm{const}$. These two distinct behaviors of $z_o$ lead to two distinct behaviors of the free stream wind velocity, which can be approximated as the average wind velocity calculated from the log-law (equation~(\ref{ulog})) evaluated at a height $H$ that is proportional to the boundary layer thickness ($H\propto\delta=\mathrm{const}$):
\begin{linenomath*}
\begin{equation}
 U_\infty\simeq\frac{u_\ast}{\kappa}\ln\frac{H}{z_o}. \label{ShearFreeStream}
\end{equation}
\end{linenomath*}
In fact, equation~(\ref{ShearFreeStream}) implies for absent (or strongly undersaturated) and saturated transport, respectively,
\begin{linenomath*}
\begin{subequations}
\begin{align}
 u_\ast&=\alpha_1U_\infty, \label{ShearFreeStream1} \\
 u_\ast&=\alpha_2(U_\infty-u_f), \label{ShearFreeStream2}
\end{align}
\end{subequations}
\end{linenomath*} 
where $\alpha_1\equiv\kappa/\ln(H/z_o)$ and $\alpha_2\equiv\kappa/\ln(H/z_f)$ are approximate constants. The dynamic saltation threshold then results from the intersection of these two relations:
\begin{linenomath*}
\begin{equation}
 u^{z_o}_t=\frac{\alpha_1\alpha_2u_f}{\alpha_2-\alpha_1}.
\end{equation}
\end{linenomath*}
That is, from measuring the relationship between $u_\ast$ and $U_\infty$ for both absent and saturated transport, we can infer $u^{z_o}_t$ from fitting equations~(\ref{ShearFreeStream1}) and (\ref{ShearFreeStream2}) to the measurements, where $\alpha_1$, $\alpha_2$, and $u_f$ are the fit parameters~\citep{Ho12}. Note that \citet{Martinetal13} identified $u_t$ in an equivalent manner from measuring the relationship between the average wind velocity at a constant large elevation and $u_\ast$. Further, note that this method exploiting properties of saturated transport does not imply that transport at the associated dynamic threshold $u^{z_o}_t$ is saturated because equation~(\ref{ShearFreeStream2}) is effectively extrapolated to vanishing transport described by equation~(\ref{ShearFreeStream1}), which neglects the transitional region that occurs near the dynamic threshold because transport is intermittent~\citep{MartinKok18}.

\section{Results} \label{Results}
\begin{figure}[htb]
 \begin{center}
  \includegraphics[width=1.0\columnwidth]{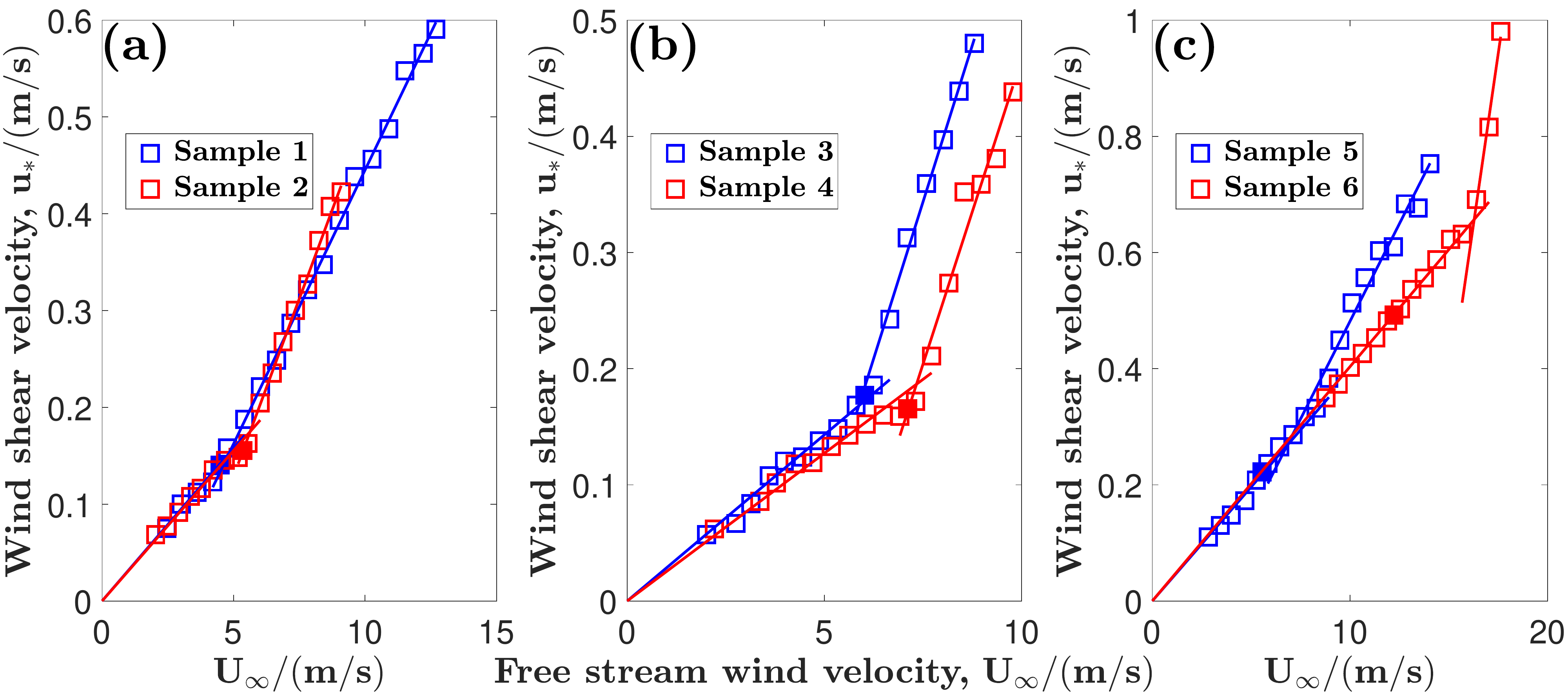}
 \end{center}
 \caption{Relationship between wind shear velocity $u_\ast$ and free stream wind velocity $U_\infty$ for a representative run for each sand sample: (a) for sand Samples~1 and 2, (b) for sand Samples~3 and 4, and (c) for sand Samples~5 and 6. The lines correspond to fits to the measurements (open squares) using equations~(\ref{ShearFreeStream1}) and (\ref{ShearFreeStream2}), the intersections of which correspond to the dynamic thresholds estimated from the roughness method ($u^{z_o}_t$). The visually measured dynamic thresholds ($u^{\mathrm{vis}}_t$) are shown as closed squares.}
\label{Measurements}
\end{figure}

For each sand sample, we carried out three or more experimental runs. Figure~\ref{Measurements} shows the measured relationship between the wind shear velocity $u_\ast$ and free stream wind velocity $U_\infty$ (open squares) the fits (lines) to the low and high measurement values using equations~(\ref{ShearFreeStream1}) and (\ref{ShearFreeStream2}), respectively, and the visually measured dynamic thresholds $u^{\mathrm{vis}}_t$ (closed squares) for a representative run for each sand sample.
The threshold values shown in Table~\ref{TableMeasurements} correspond to the values of $u^{\mathrm{vis}}_t$ and $u^{z_o}_t$ averaged over all experimental runs (number $N_r$) and their 95\% confidence intervals, which we estimated from assuming that the standard error $\sqrt{\frac{1}{(N_r-1)N_r}\sum_i^{N_r}(u_t^i-\overline{u}_t)^2}$ is the standard deviation of a Student's $t$ distribution with $N_r-1$ degrees of freedom.
\begin{table}[htb]
    \begin{tabular}{ | l | c | c | }
    \hline
     Sand sample & Visual threshold, $u^{\mathrm{vis}}_t/(\text{cm/s})$ & Roughness threshold, $u^{z_o}_t/(\text{cm/s})$ \\ 
		\hline
		 Sample~1 & $14.6\pm1.3$ & $16.2\pm2.2$ \\
		 Sample~2 & $15.4\pm2.7$ & $17.4\pm1.1$ \\
		 Sample~3 & $17.2\pm1.0$ & $17.6\pm2.0$ \\
		 Sample~4 & $16.3\pm6.0$ & $18.7\pm3.3$ \\
		 Sample~5 & $23.2\pm2.1$ & $27.4\pm1.8$ \\
		 Sample~6 & $49.3\pm1.9$ & $65.2\pm2.3$ \\
		 \hline
    \end{tabular}
		\caption{Threshold Measurements and 95\% Confidence Intervals.}
		\label{TableMeasurements}
\end{table}

\begin{figure}[htb]
 \begin{center}
  \includegraphics[width=1.0\columnwidth]{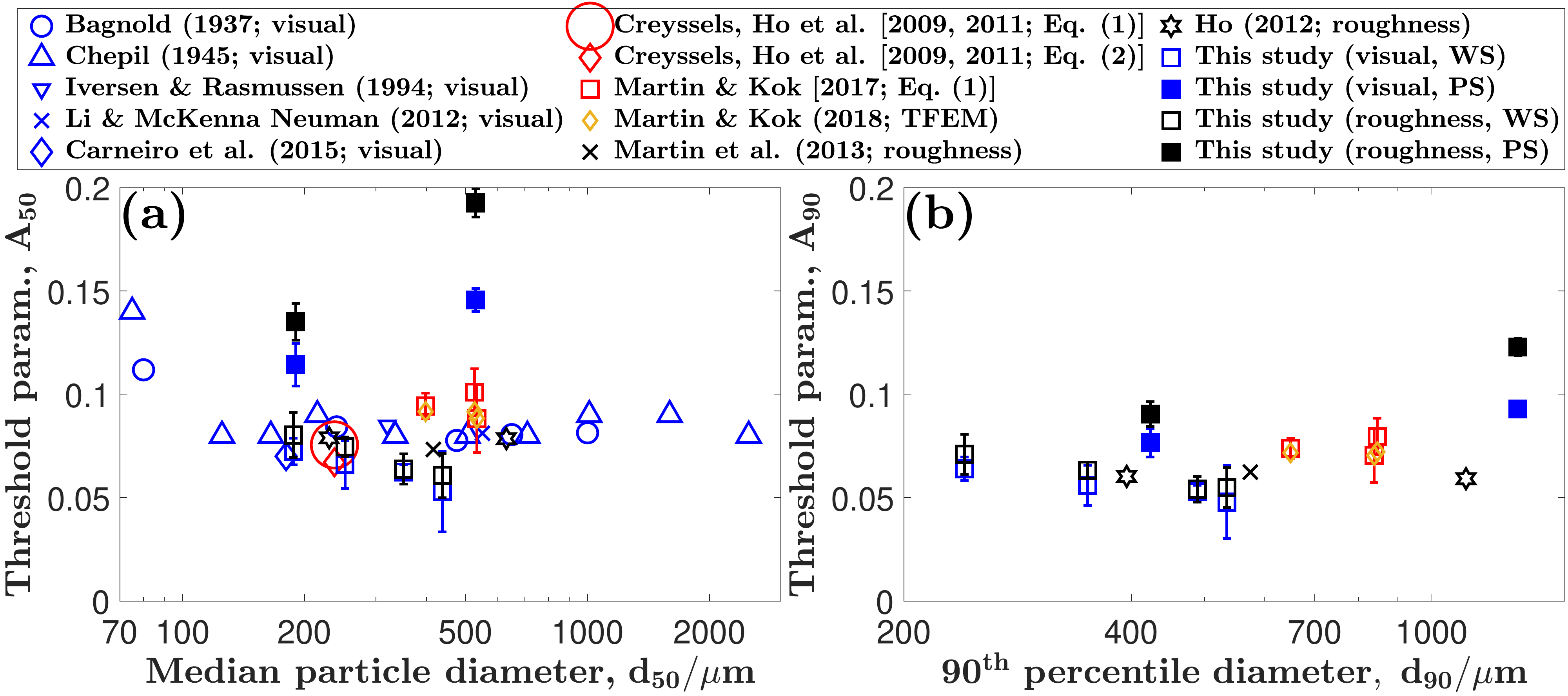}
 \end{center}
 \caption{Measurements of the threshold parameters (symbols) and 95\% confidence intervals (if known). (a) Threshold parameter $A_{50}=u_t/\sqrt{(\rho_{\mathrm{p}}/\rho_{\mathrm{a}}-1)gd_{50}}$ measured in the present (WS~=~well-sorted sand samples, PS~=~poorly sorted sand samples) and previous studies~\citep{Bagnold37,Chepil45,IversenRasmussen94,Creysselsetal09,Hoetal11,Ho12,LiMcKennaNeumann12,Martinetal13,Carneiroetal15,MartinKok17,MartinKok18} versus median particle diameter $d_{50}$ (if unknown, $d_{50}$ is assumed to be equal to mean diameter). (b) Threshold parameter $A_{90}=u_t/\sqrt{(\rho_{\mathrm{p}}/\rho_{\mathrm{a}}-1)gd_{90}}$ versus the 90th percentile particle diameter $d_{90}$ (if known). The confidence intervals are indicated by error bars. If a symbol does not consist of an error bar, the confidence interval is either indicated by the symbol size (symbols corresponding to \citep{Creysselsetal09,Hoetal11,MartinKok18}) or unknown (all other symbols without error bars). The colors encode the experimental method used to determine $u_t$: blue~=~visual method; red~=~extrapolation method (weighted least-squares) using either equation~(\ref{QCreyssels}) or equation~(\ref{QPahtz}) for the extrapolation; brown~=~Time Frequency Equivalence Method (TFEM); black~=~roughness method.}
\label{Threshold}
\end{figure}

Because the particle density $\rho_{\mathrm{p}}$ of the sand particles used for previous measurements of $u_t$ varied considerably (\citet{Chepil45} reported values between $1650$~kg/m$^3$ and $2580$~kg/m$^3$), we nondimensionalize our dynamic threshold measurements. We do so in two different ways, yielding two different \textit{threshold parameters}: the threshold parameter $A_{50}\equiv u_t/\sqrt{(\rho_{\mathrm{p}}/\rho_{\mathrm{a}}-1)gd_{50}}=\sqrt{\Theta_t}$ with respect to the median particle diameter $d_{50}$ and the threshold parameter $A_{90}\equiv u_t/\sqrt{(\rho_{\mathrm{p}}/\rho_{\mathrm{a}}-1)gd_{90}}$ with respect to the $90^{\mathrm{th}}$ percentile particle diameter $d_{90}$. Figure~\ref{Threshold}a shows the measured values of $A_{50}$ and their 95\% confidence intervals (if known) as a function of $d_{50}$ and compares them with our estimates from the weighted least-squares fits to the $Q_\ast(\Theta)$-data sets by \citet{Creysselsetal09} and \citet{Hoetal11} (Figure~\ref{ExtrapolationProcedure}), with previous wind tunnel measurements~\citep{Bagnold37,Chepil45,IversenRasmussen94,Ho12,LiMcKennaNeumann12,Carneiroetal15}, and with the field measurements by \citet{MartinKok17,MartinKok18} (who estimated $u_t$ using a weighted least-squares extrapolation method and a refined Time Frequency Equivalence Method (TFEM)~\citep{Wiggsetal04}) and \citet{Martinetal13}. 
Figure~\ref{Threshold}b shows the same as Figure~\ref{Threshold}a with respect to $d_{90}$ for those experiments for which this value is known.

\section{Discussion}
We now discuss two main observations that can be made from Figures~\ref{Measurements} and \ref{Threshold}: The visual method yields smaller thresholds than the roughness method, especially for our poorly sorted sand beds (section~\ref{VisualvsRoughness}), and both the visual and roughness method yield much larger thresholds for poorly sorted sands than for well-sorted sands with similar median particle diameter (section~\ref{SortingEffect}).

\subsection{More Than One Dynamic Threshold} \label{VisualvsRoughness}
For all our tested sand beds, the visually estimated dynamic threshold $u^{\mathrm{vis}}_t$ is smaller than the one estimated from the roughness method ($u^{z_o}_t$). While the difference between both estimations is small for our well-sorted sands ($u^{z_o}_t/u^{\mathrm{vis}}_t\simeq1.1$), which may well be attributed to a systematic underestimation of $u^{\mathrm{vis}}_t$ (section~\ref{VisualMethod}), it is quite significant for our poorly sorted sands ($u^{z_o}_t/u^{\mathrm{vis}}_t\simeq1.2$ for Sample~5 and $u^{z_o}_t/u^{\mathrm{vis}}_t\simeq1.3$ for Sample~6). This is a curious finding because the current consensus is that there is a single threshold associated with the cessation of intermittent saltation transport for polydisperse sand beds~\citep{MartinKok18,MartinKok19}.

As we described in section~\ref{RoughnessMethod}, the roughness method exploits that saltation transport can only saturate when there is a strong suppression of near-bed wind speeds from the particle-flow feedback, which leads to a substantial increase of the surface roughness $z_o$. The fact that there is no such increase at $u^{\mathrm{vis}}_t$ (e.g., the closed square in Figure~\ref{Measurements} corresponding to Sample~6 lies well within the region where $u_\ast\propto U_\infty$) thus likely implies that transport is strongly undersaturated. This does not mean that saltation transport is unsteady but rather that there is a different equilibrium between bed sediment entrainment and deposition when compared with the saturated state. While in the latter case, the equilibrium is probably caused by the particle-flow feedback spiking the deposition rate (section~\ref{RoughnessMethod}), we believe that, in the former case, it is caused by the limited availability of erodible fine bed surface particles because of armoring by coarse particles. In other words, saturated transport is deposition-limited, whereas the here found undersaturated transport is erosion-limited. Note that, even though armoring may cause considerable spatial and temporal variability of saltation transport, this potential variability is unlikely to have affected our wind speed measurements because of the near absence of the particle-flow feedback for the undersaturated conditions discussed here (i.e., $u^{\mathrm{vis}}_t<u_\ast<u^{z_o}_t$ for Samples~5 and 6).

The hypothesis that armoring is responsible for saltation transport being undersaturated is supported by our observation that, for Samples~5 and 6, only relatively fine particles were saltating for $u^{\mathrm{vis}}_t<u_\ast<u^{z_o}_t$, whereas relatively coarse particles (found near the crest of the ripples) crept along the surface, consistent with the ripples being megaripples~\citep{Katraetal14,Lammeletal18}. This observation means that $u^{\mathrm{vis}}_t$ measures the dynamic saltation threshold of a subset of relatively fine particles, whereas $u^{z_o}_t$ measures the dynamic threshold of the entire ensemble of particles, which indicates that a size-selective process controls $u_t$. Hence, like \citet{PahtzDuran18a}, we hypothesize that $u_t$ is the minimal wind shear velocity that is needed to compensate energy losses of rebounding particles during particle trajectories, which is a size-selective process because larger particles are accelerated less strongly during their hops. Note that the finding of size selectivity does not contradict the field measurements by \citet{MartinKok19}, which showed that the size distribution of particles in saltation is relatively insensitive to the wind shear velocity $u_\ast$, because the sand beds at these authors' field sites were considerably better sorted ($d_{90}/d_{50}\approx1.6$) than our Samples~5 ($d_{90}/d_{50}\approx2.2$) and 6 ($d_{90}/d_{50}\approx2.5$) and thus likely did not exhibit a significant difference between $u^{\mathrm{vis}}_t$ and $u^{z_o}_t$ (like our Samples~1-4).

In summary, for poorly sorted sand beds, there may be three distinct dynamic thresholds: a threshold associated with the cessation of intermittent saltation transport of relatively fine particles ($u^{\mathrm{vis}}_t$), a larger threshold associated with the cessation of intermittent saltation transport of the entire ensemble of particles ($u^{z_o}_t$), and an even larger threshold below which continuous saltation transport becomes intermittent (there is an ongoing controversy about whether this threshold is associated with splash entrainment or aerodynamic entrainment~\citep{MartinKok18,Pahtz18}).

\subsection{Much Larger Threshold for Poorly Sorted Than for Well-Sorted Sand} \label{SortingEffect}
Samples~1 and 5 and Samples~4 and 6 exhibit a similar median particle diameter (Figure~\ref{WindTunnel}b). Yet both the dynamic thresholds estimated from the visual ($u^{\mathrm{vis}}_t$) and roughness method ($u^{z_o}_t$) differ greatly between these samples (Table~\ref{TableMeasurements} and \ref{Threshold}a). Much of this divergence seems to be caused by the presence of relatively very coarse particles in the bed as even a rescaling based on $d_{90}$ rather than $d_{50}$ cannot fully explain the spread between existing measurements (Figure~\ref{Threshold}b).

\subsubsection{Visual Threshold}
We propose that $u^{\mathrm{vis}}_t$ increases with particle size heterogeneity at least partly because of hiding effects in heterogeneous sand beds: Relatively fine particles tend to be surrounded by coarser ones, and their \textit{protrusion} (i.e., the particle height above surrounding sediment) is thus smaller than on average, whereas relatively coarse particles tend to have a larger-than-average protrusion. Importantly, \citet{Yageretal18} showed that a particle's protrusion does affect not only the driving flow forces acting on this particle but also its ability to resist entrainment: Smaller protrusion is associated with larger resisting forces. It seems conceivable that a particle colliding with the bed surface at a location associated with a small protrusion at the moment of impact also experiences larger forces resisting its ability to rebound~\citep{Yageretal18}. That is, assuming that $u^{\mathrm{vis}}_t$ is associated with sustained rebounds of a subset of relatively fine particles of the entire particle ensemble (our hypothesis in section~\ref{VisualvsRoughness}), its value should increase with sand size heterogeneity, as observed.

A second effect that potentially leads to an increase of $u^{\mathrm{vis}}_t$ is the armoring of fine particles by coarse particles gathering at the ripple crests, which makes it more difficult to entrain bed sediment. Even if $u^{\mathrm{vis}}_t$ is associated with sustained rebounds of fine particles rather than their entrainment by splash or aerodynamic forces, as we hypothesized in section~\ref{VisualvsRoughness}, armoring should lead to an increase of its value because particles lose a larger fraction of their kinetic energy on average when rebounding with a bed made of larger particles than when rebounding with a bed made of particles of the same size~\citep{Lammeletal17}. Note that this effect is physically similar, if not equivalent, to the protrusion effect described above because the larger energy loss is caused by the larger probability of particle rebounds at locations with large protrusion~\citep{Lammeletal17} as they are associated with larger forces resisting the rebounds~\citep{Yageretal18}.

\subsubsection{Roughness Threshold}
Assuming that $u^{z_o}_t$ is associated with sustained rebounds of the entire ensemble of particles (our hypothesis in section~\ref{VisualvsRoughness}), its value should be controlled by the ability of the flow to sustain rebounds of relatively coarse particles. Because such particles have a harder job to maintain their bouncing motion than the median particle, since they experience less fluid drag acceleration during their hops, this assumption automatically explains why $u^{z_o}_t$ is larger for heterogeneous sand beds than for homogeneous ones. Furthermore, we would like to emphasize that the ability of a certain particle class to rebound is probably strongly tied to its ability to eject particles of the same size class via splash~\citep[][section~4.2.1]{PahtzDuran18a}. Hence, if splash-sustained transport required that all size classes are equally susceptible to splash entrainment (as it likely does~\citep{MartinKok19}), this assumption would also explain why the relationship between $u_\ast$ and $U_\infty$ tends to be described by equation~(\ref{ShearFreeStream2}) at shear velocities $u_\ast$ that are not too far above $u^{z_o}_t$, as splash-sustained transport is the origin of the strong particle-flow feedback causing the roughness increase described by equation~(\ref{ShearFreeStream2}) (section~\ref{RoughnessMethod}). However, note that the shift from equation~(\ref{ShearFreeStream1}) to equation~(\ref{ShearFreeStream2}) does not always happen immediately (e.g., there is an obvious transitional region for Sample~5 in Figure~\ref{Measurements}), which is also consistent with the rebound hypothesis (in the sustained-rebound picture, the splash entrainment threshold is always larger than the rebound threshold~\citep{PahtzDuran18a}).

\section{Conclusions}
In this study, we have measured in a wind tunnel, or determined from existing experimental data sets, the dynamic saltation threshold $u_t$ by three different means: by extrapolating paired measurements of the wind shear velocity $u_\ast$ and transport rate to vanishing transport, by decrementing $u_\ast$ and visually estimating its value when transport stops, and by exploiting a regime shift in the behavior of the surface roughness caused by momentum transfer from the wind to the saltating particles. All three methods yield threshold values that are consistent with each other for sufficiently well-sorted sand beds provided that the extrapolation method takes measurement uncertainties into account (Figure~\ref{Threshold}a). However, there is a strongly increasing trend of $u_t$ with sand size heterogeneity that even a rescaling based on the 90th percentile particle diameter $d_{90}$ (replacing the median diameter $d_{50}$) cannot fully capture (Figure~\ref{Threshold}b), which suggests that relatively very coarse particles ($d>d_{90}$) have a considerable control on the dynamic threshold. For example, $u_t$ estimated from the roughness method differs by a factor of $3.5$ for two of our tested sands (Samples~4 and 6 in Table~\ref{TableMeasurements}) despite having a similar $d_{50}$ (Figure~\ref{WindTunnel}b). We have offered an explanation for this remarkable finding based on hiding effects and sustained particle rebounds in heterogeneous sand beds (section~\ref{SortingEffect}). Interestingly, a predominant effect of relatively very coarse particles on the mobility of the bed was previously reported also for water-driven sediment transport~\citep{MacKenzieEaton17}, which led \citet{MacKenzieetal18} to challenge the long-standing assumption that $d_{50}$ is the best choice for the characteristic size of bed particles. Our study challenges this assumption also for wind-driven sediment transport.

Furthermore, sufficiently heterogeneous sand beds exhibit more than one dynamic threshold (for relatively uniform samples, our results are inconclusive), likely because of size-selective processes (section~\ref{VisualvsRoughness}). This is not a trivial finding because $u_t$ is often seen as a quantity that describes the entire ensemble rather than a subset of bed particles~\citep[e.g.,][]{ClaudinAndreotti06} and because \citet{MartinKok19} recently reported for moderately heterogeneous sand beds in the field that fine and coarse particles are equally susceptible to participate in saltation transport (i.e., there is only a single dynamic threshold). In combination, these authors' and our findings hint at the possibility that, in order for fine and coarse grains to have different susceptibility to saltate, the level of heterogeneity and/or type (e.g., unimodal versus bimodal) of the size distribution of bed surface particles are important, both of which change with ongoing aeolian transport~\citep{Lammeletal18}.

The large effects of particle size heterogeneity found in this study have important implications for quantitative predictions of different kinds of geophysical processes. For example, given that the particle size distribution in the field can vary from very heterogeneous (e.g., our sand from Tengger Desert) to relatively uniform~\citep[e.g.,][]{Martinetal13}, basing the dynamic threshold only on the median diameter in theoretical models (e.g., dust aerosol emission schemes in climate models~\citep{Hausteinetal15} or models of extraterrestrial transport~\citep{Telferetal18}) may lead to fundamentally wrong predictions. Finally, one may also wonder to what degree sand size heterogeneity has contributed to the generally large mismatch between theoretical predictions (uniform sand) and laboratory and field measurements (often very heterogeneous sand) of aeolian sand transport, which is currently attributed mostly to temporal and spatial variability in the field~\citep{Barchynetal14}. In particular, the here reported size selectivity of the dynamic threshold may leave a signature in aeolian transport laws and thus lead to qualitative deviations between theory and field beyond the mere quantitative influences of the threshold value and scaling constants.

Finally, it is worth mentioning that the insights into the behavior of the dynamic saltation threshold gained from this study may also help to better understand the static threshold of saltation transport. In fact, based on a theoretical analysis and a compilation of wind tunnel measurements, \citet{Pahtzetal18} recently showed that the static saltation threshold depends on the thickness of the turbulent boundary layer and argued that, for field conditions (very thick boundary layer), an episodic short-lived rolling motion of isolated particles may be initiated below the dynamic saltation threshold. Such episodic rolling can evolve into fully developed saltation transport only if the shear velocity is above the dynamic saltation threshold, which led \citet{Pahtzetal18} to propose that, for many field conditions (including on Mars), the static saltation threshold may actually be controlled by dynamic mechanisms and close to the dynamic saltation threshold.

\acknowledgments
The flow velocity profiles for each sand sample for each run can be found on the Zenodo data repository at \url{https://zenodo.org/record/2550975#.XE67ofZuLSE}. We thank Klaus Kroy and Katharina Tholen for interesting discussions related to this study. We thank Keld R{\o}mer Rasmussen and various anonymous reviewers for their constructive reviews. We acknowledge support from grant National Natural Science Foundation of China (No.~11750410687, No.~11772143, and No.~11672267) and partial support from the Fundamental Research Funds for the Central Universities (2017XZZX001-02A).


\begin{thebibliography}{55}
\providecommand{\natexlab}[1]{#1}
\expandafter\ifx\csname urlstyle\endcsname\relax
  \providecommand{\doi}[1]{doi:\discretionary{}{}{}#1}\else
  \providecommand{\doi}{doi:\discretionary{}{}{}\begingroup
  \urlstyle{rm}\Url}\fi

\bibitem[{\textit{Bagnold}(1936)}]{Bagnold36}
Bagnold, R.~A. (1936), The movement of desert sand, \textit{Proceedings of the
  Royal Society London Series A}, \textit{157}, 594--620.

\bibitem[{\textit{Bagnold}(1937)}]{Bagnold37}
Bagnold, R.~A. (1937), The transport of sand by wind, \textit{The Geographical
  Journal}, \textit{89}(5), 409--438, \doi{10.2307/1786411}.

\bibitem[{\textit{Bagnold}(1941)}]{Bagnold41}
Bagnold, R.~A. (1941), \textit{The Physics of Blown Sand and Desert Dunes},
  Methuen, New York.

\bibitem[{\textit{Barchyn and Hugenholtz}(2011)}]{BarchynHugenholtz11}
Barchyn, T.~E., and C.~H. Hugenholtz (2011), Comparison of four methods to
  calculate aeolian sediment transport threshold from field data: Implications
  for transport prediction and discussion of method evolution,
  \textit{Geomorphology}, \textit{129}, 190--203,
  \doi{10.1016/j.geomorph.2011.01.022}.

\bibitem[{\textit{Barchyn et~al.}(2014)\textit{Barchyn, Martin, Kok, and
  Hugenholtz}}]{Barchynetal14}
Barchyn, T.~E., R.~L. Martin, J.~F. Kok, and C.~H. Hugenholtz (2014),
  Fundamental mismatches between measurements and models in aeolian sediment
  transport prediction: The role of small-scale variability, \textit{Aeolian
  Research}, \textit{15}, 245--251, \doi{10.1016/j.aeolia.2014.07.002}.

\bibitem[{\textit{Bourke et~al.}(2010)\textit{Bourke, Lancaster, Fenton,
  Parteli, Zimbelman, and Radebaugh}}]{Bourkeetal10}
Bourke, M.~C., N.~Lancaster, L.~K. Fenton, E.~J.~R. Parteli, J.~R. Zimbelman,
  and J.~Radebaugh (2010), Extraterrestrial dunes: An introduction to the
  special issue on planetary dune systems, \textit{Geomorphology},
  \textit{121}, 1--14, \doi{10.1016/j.geomorph.2010.04.007}.

\bibitem[{\textit{Burr et~al.}(2015)\textit{Burr, Bridges, Marshall, Smith,
  White, and Emery}}]{Burretal15}
Burr, D.~M., N.~T. Bridges, J.~R. Marshall, J.~K. Smith, B.~R. White, and J.~P.
  Emery (2015), Higher-than-predicted saltation threshold wind speeds on
  {Titan}, \textit{Nature}, \textit{517}(7532), 60--63,
  \doi{10.1038/nature14088}.

\bibitem[{\textit{Carneiro et~al.}(2013)\textit{Carneiro, Ara\'ujo, P\"ahtz,
  and Herrmann}}]{Carneiroetal13}
Carneiro, M.~V., N.~A.~M. Ara\'ujo, T.~P\"ahtz, and H.~J. Herrmann (2013),
  Midair collisions enhance saltation, \textit{Physical Review Letters},
  \textit{111}(5), 058,001, \doi{10.1103/PhysRevLett.111.058001}.

\bibitem[{\textit{Carneiro et~al.}(2015)\textit{Carneiro, Rasmussen, and
  Herrmann}}]{Carneiroetal15}
Carneiro, M.~V., K.~R. Rasmussen, and H.~J. Herrmann (2015), Bursts in
  discontinuous aeolian saltation, \textit{Scientific Reports}, \textit{5},
  11,109, \doi{10.1038/srep11109}.

\bibitem[{\textit{Chepil}(1945)}]{Chepil45}
Chepil, W.~S. (1945), Dynamics of wind erosion: {II. Initiation} of soil
  movement, \textit{Soil Science}, \textit{60}(5), 397--411,
  \doi{10.1097/00010694-194511000-00005}.

\bibitem[{\textit{Claudin and Andreotti}(2006)}]{ClaudinAndreotti06}
Claudin, P., and B.~Andreotti (2006), A scaling law for aeolian dunes on
  {Mars}, {Venus}, {Earth}, and for subaqueous ripples, \textit{Earth and
  Planetary Science Letters}, \textit{252}, 30--44,
  \doi{10.1016/j.epsl.2006.09.004}.

\bibitem[{\textit{Creyssels et~al.}(2009)\textit{Creyssels, Dupont, {Ould El
  Moctar}, Valance, Cantat, Jenkins, Pasini, and Rasmussen}}]{Creysselsetal09}
Creyssels, M., P.~Dupont, A.~{Ould El Moctar}, A.~Valance, I.~Cantat, J.~T.
  Jenkins, J.~M. Pasini, and K.~R. Rasmussen (2009), Saltating particles in a
  turbulent boundary layer: experiment and theory, \textit{Journal of Fluid
  Mechanics}, \textit{625}, 47--74, \doi{10.1017/S0022112008005491}.

\bibitem[{\textit{{de Vet} et~al.}(2014)\textit{{de Vet}, Merrison,
  Mittelmeijer-Hazeleger, {van Loon}, and Cammeraat}}]{deVetetal14}
{de Vet}, S.~J., J.~P. Merrison, M.~C. Mittelmeijer-Hazeleger, E.~E. {van
  Loon}, and L.~H. Cammeraat (2014), Effects of rolling on wind-induced
  detachment thresholds of volcanic glass on {Mars}, \textit{Planetary and
  Space Science}, \textit{103}, 205--218, \doi{10.1016/j.pss.2014.07.012}.

\bibitem[{\textit{Dur\'an et~al.}(2011)\textit{Dur\'an, Claudin, and
  Andreotti}}]{Duranetal11}
Dur\'an, O., P.~Claudin, and B.~Andreotti (2011), On aeolian transport:
  Grain-scale interactions, dynamical mechanisms and scaling laws,
  \textit{Aeolian Research}, \textit{3}, 243--270,
  \doi{10.1016/j.aeolia.2011.07.006}.

\bibitem[{\textit{Dur\'an et~al.}(2012)\textit{Dur\'an, Andreotti, and
  Claudin}}]{Duranetal12}
Dur\'an, O., B.~Andreotti, and P.~Claudin (2012), Numerical simulation of
  turbulent sediment transport, from bed load to saltation, \textit{Physics of
  Fluids}, \textit{24}, 103,306, \doi{10.1063/1.4757662}.

\bibitem[{\textit{Dur\'an et~al.}(2014{\natexlab{a}})\textit{Dur\'an,
  Andreotti, and Claudin}}]{Duranetal14a}
Dur\'an, O., B.~Andreotti, and P.~Claudin (2014{\natexlab{a}}), Turbulent and
  viscous sediment transport - a numerical study, \textit{Advances in
  Geosciences}, \textit{37}, 73--80, \doi{10.5194/adgeo-37-73-2014}.

\bibitem[{\textit{Dur\'an et~al.}(2014{\natexlab{b}})\textit{Dur\'an, Claudin,
  and Andreotti}}]{Duranetal14b}
Dur\'an, O., P.~Claudin, and B.~Andreotti (2014{\natexlab{b}}), Direct
  numerical simulations of aeolian sand ripples, \textit{Proceedings of the
  National Academy of Science}, \textit{111}(44), 15,665--15,668,
  \doi{10.1073/pnas.1413058111}.

\bibitem[{\textit{Gillette et~al.}(1980)\textit{Gillette, Adams, Endo, Smith,
  and Kihl}}]{Gilletteetal80}
Gillette, D.~A., J.~Adams, A.~Endo, D.~Smith, and R.~Kihl (1980), Threshold
  velocities for input of soil particles into the air by desert soils,
  \textit{Journal of Geophysical Research}, \textit{85}(C10), 5621--5630,
  \doi{10.1029/JC085iC10p05621}.

\bibitem[{\textit{Haustein et~al.}(2015)\textit{Haustein, Washington, King,
  Wiggs, Thomas, Eckardt, Bryant, and Menut}}]{Hausteinetal15}
Haustein, K., R.~Washington, J.~King, G.~Wiggs, D.~S.~G. Thomas, F.~D. Eckardt,
  R.~G. Bryant, and L.~Menut (2015), Testing the performance of
  state-of-the-art dust emission schemes using do4models field data,
  \textit{Geoscientific Model Development}, \textit{8}, 341--362,
  \doi{10.5194/gmd-8-341-2015}.

\bibitem[{\textit{Ho}(2012)}]{Ho12}
Ho, T.~D. (2012), Etude exp\'erimentale du transport de particules dans une
  couche limite turbulente, Ph.D. thesis, University of Rennes, Rennes, France.

\bibitem[{\textit{Ho et~al.}(2011)\textit{Ho, Valance, Dupont, and {Ould El
  Moctar}}}]{Hoetal11}
Ho, T.~D., A.~Valance, P.~Dupont, and A.~{Ould El Moctar} (2011), Scaling laws
  in aeolian sand transport, \textit{Physical Review Letters}, \textit{106},
  094,501, \doi{10.1103/PhysRevLett.106.094501}.

\bibitem[{\textit{Iversen et~al.}(1987)\textit{Iversen, Greeley, Marshall, and
  Pollack}}]{Iversenetal87}
Iversen, J., R.~Greeley, J.~R. Marshall, and J.~B. Pollack (1987), Aeolian
  saltation threshold: the effect of density ratio, \textit{Sedimentology},
  \textit{34}, 699--706, \doi{10.1111/j.1365-3091.1987.tb00795.x}.

\bibitem[{\textit{Iversen and Rasmussen}(1994)}]{IversenRasmussen94}
Iversen, J.~D., and K.~R. Rasmussen (1994), The effect of surface slope on
  saltation threshold, \textit{Sedimentology}, \textit{41}, 721--728,
  \doi{10.1111/j.1365-3091.1994.tb01419.x}.

\bibitem[{\textit{Katra et~al.}(2014)\textit{Katra, H, and Kok}}]{Katraetal14}
Katra, I., Y.~H, and J.~F. Kok (2014), Mechanisms limiting the growth of
  aeolian megaripples, \textit{Geophysical Research Letters}, \textit{41},
  858--865, \doi{10.1002/2013GL058665}.

\bibitem[{\textit{Kok et~al.}(2012)\textit{Kok, Parteli, Michaels, and
  Karam}}]{Koketal12}
Kok, J.~F., E.~J.~R. Parteli, T.~I. Michaels, and D.~B. Karam (2012), The
  physics of wind-blown sand and dust, \textit{Reports on Progress in Physics},
  \textit{75}, 106,901, \doi{10.1088/0034-4885/75/10/106901}.

\bibitem[{\textit{Kok et~al.}(2014{\natexlab{a}})\textit{Kok, Mahowald,
  Fratini, Gillies, Ishizuka, Leys, Mikami, Park, Park, Pelt, and
  Zobeck}}]{Koketal14a}
Kok, J.~F., N.~M. Mahowald, G.~Fratini, J.~A. Gillies, M.~Ishizuka, J.~F. Leys,
  M.~Mikami, M.-S. Park, S.-U. Park, R.~S.~V. Pelt, and T.~M. Zobeck
  (2014{\natexlab{a}}), An improved dust emission model - part 1: Model
  description and comparison against measurements, \textit{Atmospheric
  Chemistry and Physics}, \textit{14}, 13,023--13,041,
  \doi{10.5194/acp-14-13023-2014}.

\bibitem[{\textit{Kok et~al.}(2014{\natexlab{b}})\textit{Kok, Albani, Mahowald,
  and Ward}}]{Koketal14b}
Kok, J.~F., S.~Albani, N.~M. Mahowald, and D.~S. Ward (2014{\natexlab{b}}), An
  improved dust emission model - part 2: Evaluation in the {Community Earth
  System Model}, with implications for the use of dust source functions,
  \textit{Atmospheric Chemistry and Physics}, \textit{14}, 13,043--13,061,
  \doi{10.5194/acp-14-13043-2014}.

\bibitem[{\textit{Kok et~al.}(2018)\textit{Kok, Ward, Mahowald, and
  Evan}}]{Koketal18}
Kok, J.~F., D.~S. Ward, N.~M. Mahowald, and A.~T. Evan (2018), Global and
  regional importance of the direct dust-climate feedback, \textit{Nature
  Communications}, \textit{9}, 241, \doi{10.1038/s41467-017-02620-y}.

\bibitem[{\textit{L\"ammel et~al.}(2017)\textit{L\"ammel, Dzikowski, Kroy,
  Oger, and Valance}}]{Lammeletal17}
L\"ammel, M., K.~Dzikowski, K.~Kroy, L.~Oger, and A.~Valance (2017),
  Grain-scale modeling and splash parametrization for aeolian sand transport,
  \textit{Physical Review E}, \textit{95}, 022,902,
  \doi{10.1103/PhysRevE.95.022902}.

\bibitem[{\textit{L\"ammel et~al.}(2018)\textit{L\"ammel, Meiwald, Yizhaq,
  Tsoar, Katra, and Kroy}}]{Lammeletal18}
L\"ammel, M., A.~Meiwald, H.~Yizhaq, H.~Tsoar, I.~Katra, and K.~Kroy (2018),
  Aeolian sand sorting and megaripple formation, \textit{Nature Physics},
  \textit{14}, 759--765, \doi{10.1038/s41567-018-0106-z}.

\bibitem[{\textit{Li and {McKenna Neumann}}(2012)}]{LiMcKennaNeumann12}
Li, B., and C.~{McKenna Neumann} (2012), Boundary-layer turbulence
  characteristics during aeolian saltation, \textit{Geophysical Research
  Letters}, \textit{39}(11), L11,402, \doi{10.1029/2012GL052234}.

\bibitem[{\textit{Li et~al.}(2014)\textit{Li, Ellis, and Sherman}}]{Lietal14}
Li, B., J.~T. Ellis, and D.~J. Sherman (2014), Estimating the impact threshold
  for wind-blown sand, \textit{Journal of Coastal Research}, \textit{70(spl)},
  627--632, \doi{10.2112/SI70-106.1}.

\bibitem[{\textit{MacKenzie and Eaton}(2017)}]{MacKenzieEaton17}
MacKenzie, L.~G., and B.~C. Eaton (2017), Large grains matter: contrasting bed
  stability and morphodynamics during two nearly identical experiments,
  \textit{Earth Surface Processes and Landforms}, \textit{42}, 1287--1295,
  \doi{10.1002/esp.4122}.

\bibitem[{\textit{MacKenzie et~al.}(2018)\textit{MacKenzie, Eaton, and
  Church}}]{MacKenzieetal18}
MacKenzie, L.~G., B.~C. Eaton, and M.~Church (2018), Breaking from the average:
  Why large grains matter in gravel-bed streams, \textit{Earth Surface
  Processes and Landforms}, \textit{43}, 3190--3196, \doi{10.1002/esp.4465}.

\bibitem[{\textit{Martin and Kok}(2017)}]{MartinKok17}
Martin, R.~L., and J.~F. Kok (2017), Wind-invariant saltation heights imply
  linear scaling of aeolian saltation flux with shear stress, \textit{Science
  Advances}, \textit{3}, e1602,569, \doi{10.1126/sciadv.1602569}.

\bibitem[{\textit{Martin and Kok}(2018)}]{MartinKok18}
Martin, R.~L., and J.~F. Kok (2018), Distinct thresholds for the initiation and
  cessation of aeolian saltation from field measurements, \textit{Journal of
  Geophysical Research: Earth Surface}, \textit{123}(7), 1546--1565,
  \doi{10.1029/2017JF004416}.

\bibitem[{\textit{Martin and Kok}(2019)}]{MartinKok19}
Martin, R.~L., and J.~F. Kok (2019), Size‐independent susceptibility to
  transport in aeolian saltation, \textit{Journal of Geophysical Research:
  Earth Surface}, \textit{124}(7), 1658--1674, \doi{10.1029/2019JF005104}.

\bibitem[{\textit{Martin et~al.}(2013)\textit{Martin, Barchyn, Hugenholtz, and
  Jerolmack}}]{Martinetal13}
Martin, R.~L., T.~E. Barchyn, C.~H. Hugenholtz, and D.~J. Jerolmack (2013),
  Timescale dependence of aeolian sand flux observations under atmospheric
  turbulence, \textit{Journal of Geophysical Research: Atmospheres},
  \textit{118}(16), 9078--9092, \doi{10.1002/jgrd.50687}.

\bibitem[{\textit{Merrison}(2012)}]{Merrison12}
Merrison, J.~P. (2012), Sand transport, erosion and granular electrification,
  \textit{Aeolian Research}, \textit{4}, 1--16.

\bibitem[{\textit{Merrison et~al.}(2007)\textit{Merrison, Gunnlaugsson,
  N{\o}rnberg, Jensen, and Rasmussen}}]{Merrisonetal07}
Merrison, J.~P., H.~P. Gunnlaugsson, P.~N{\o}rnberg, A.~E. Jensen, and
  K.~Rasmussen (2007), Determination of the wind induced detachment threshold
  for granular material on {Mars} using wind tunnel simulations,
  \textit{Icarus}, \textit{191}, 568--580, \doi{10.1016/j.icarus.2007.04.035}.

\bibitem[{\textit{Nickling}(1988)}]{Nickling88}
Nickling, W.~G. (1988), The initiation of particle movement by wind,
  \textit{Sedimentology}, \textit{35}(3), 499--511,
  \doi{10.1111/j.1365-3091.1988.tb01000.x}.

\bibitem[{\textit{P\"ahtz}(2018)}]{Pahtz18}
P\"ahtz, T. (2018), Comment on ``{Distinct} thresholds for the initiation and
  cessation of aeolian saltation from field measurements'' by {Raleigh L.
  Martin} and {Jasper F. Kok}: Alternative interpretation of measured
  thresholds as two distinct cessation thresholds, \textit{Journal of
  Geophysical Research: Earth Surface}, \textit{123}(12), 3388--3391,
  \doi{10.1029/2018JF004824}.

\bibitem[{\textit{P\"ahtz and Dur\'an}(2017)}]{PahtzDuran17}
P\"ahtz, T., and O.~Dur\'an (2017), Fluid forces or impacts: What governs the
  entrainment of soil particles in sediment transport mediated by a {Newtonian}
  fluid?, \textit{Physical Review Fluids}, \textit{2}(7), 074,303,
  \doi{10.1103/PhysRevFluids.2.074303}.

\bibitem[{\textit{P\"ahtz and Dur\'an}(2018{\natexlab{a}})}]{PahtzDuran18a}
P\"ahtz, T., and O.~Dur\'an (2018{\natexlab{a}}), The cessation threshold of
  nonsuspended sediment transport across aeolian and fluvial environments,
  \textit{Journal of Geophysical Research: Earth Surface}, \textit{123}(8),
  1638--1666, \doi{10.1029/2017JF004580}.

\bibitem[{\textit{P\"ahtz and Dur\'an}(2018{\natexlab{b}})}]{PahtzDuran18b}
P\"ahtz, T., and O.~Dur\'an (2018{\natexlab{b}}), Universal friction law at
  granular solid-gas transition explains scaling of sediment transport load
  with excess fluid shear stress, \textit{Physical Review Fluids},
  \textit{3}(10), 104,302, \doi{10.1103/PhysRevFluids.3.104302}.

\bibitem[{\textit{P\"ahtz et~al.}(2018)\textit{P\"ahtz, Valyrakis, Zhao, and
  Li}}]{Pahtzetal18}
P\"ahtz, T., M.~Valyrakis, X.~H. Zhao, and Z.~S. Li (2018), The critical role
  of the boundary layer thickness for the initiation of aeolian sediment
  transport, \textit{Geosciences}, \textit{8}(9), 314,
  \doi{10.3390/geosciences8090314}.

\bibitem[{\textit{Raffaele et~al.}(2016)\textit{Raffaele, Bruno, Pellerey, and
  Preziosi}}]{Raffaeleetal16}
Raffaele, L., L.~Bruno, F.~Pellerey, and L.~Preziosi (2016), Windblown sand
  saltation: A statistical approach to fluid threshold shear velocity,
  \textit{Aeolian Research}, \textit{23}, 79--91,
  \doi{10.1016/j.aeolia.2016.10.002}.

\bibitem[{\textit{Rasmussen et~al.}(2015)\textit{Rasmussen, Valance, and
  Merrison}}]{Rasmussenetal15}
Rasmussen, K.~R., A.~Valance, and J.~Merrison (2015), Laboratory studies of
  aeolian sediment transport processes on planetary surfaces,
  \textit{Geomorphology}, \textit{244}, 74--94,
  \doi{10.1016/j.geomorph.2015.03.041}.

\bibitem[{\textit{Selmani et~al.}(2018)\textit{Selmani, Valance, {Ould El
  Moctar}, Dupont, and Zegadi}}]{Selmanietal18}
Selmani, H., A.~Valance, A.~{Ould El Moctar}, P.~Dupont, and R.~Zegadi (2018),
  Aeolian sand transport in out-of-equilibrium regimes, \textit{Geophysical
  Research Letters}, \textit{45}, 1838--1844, \doi{10.1002/2017GL076937}.

\bibitem[{\textit{Telfer et~al.}(2018)\textit{Telfer, Parteli, Radebaugh,
  Beyer, Bertrand, Forget, Nimmo, Grundy, Moore, Stern, Spencer, Lauer, Earle,
  Binzel, Weaver, Olkin, Young, Ennico, Runyon, and {The New Horizons Geology,
  Geophysics and Imaging Science Theme Team}}}]{Telferetal18}
Telfer, M.~W., E.~J.~R. Parteli, J.~Radebaugh, R.~A. Beyer, T.~Bertrand,
  F.~Forget, F.~Nimmo, W.~M. Grundy, J.~M. Moore, S.~A. Stern, J.~Spencer,
  T.~R. Lauer, A.~M. Earle, R.~P. Binzel, H.~A. Weaver, C.~B. Olkin, L.~A.
  Young, K.~Ennico, K.~Runyon, and {The New Horizons Geology, Geophysics and
  Imaging Science Theme Team} (2018), Dunes on {Pluto}, \textit{Science},
  \textit{360}, 992--997, \doi{10.1126/science.aao2975}.

\bibitem[{\textit{Valance et~al.}(2015)\textit{Valance, Rasmussen, {Ould El
  Moctar}, and Dupont}}]{Valanceetal15}
Valance, A., K.~R. Rasmussen, A.~{Ould El Moctar}, and P.~Dupont (2015), The
  physics of aeolian sand transport, \textit{Comptes Rendus Physique},
  \textit{16}, 105--117, \doi{10.1016/j.crhy.2015.01.006}.

\bibitem[{\textit{Walter et~al.}(2014)\textit{Walter, Horender, Voegeli, and
  Lehning}}]{Walteretal14}
Walter, B., S.~Horender, C.~Voegeli, and M.~Lehning (2014), Experimental
  assessment of {Owen's} second hypothesis on surface shear stress induced by a
  fluid during sediment saltation, \textit{Geophysical Research Letters},
  \textit{41}, 6298--6305, \doi{10.1002/2014GL061069}.

\bibitem[{\textit{Wiggs et~al.}(2004)\textit{Wiggs, Atherton, and
  Baird}}]{Wiggsetal04}
Wiggs, G. F.~S., R.~J. Atherton, and A.~J. Baird (2004), Thresholds of aeolian
  sand transport: establishing suitable values, \textit{Sedimentology},
  \textit{51}, 95--108, \doi{10.1046/j.1365-3091.2003.0061}.

\bibitem[{\textit{Yager et~al.}(2018)\textit{Yager, Schmeeckle, and
  Badoux}}]{Yageretal18}
Yager, E.~M., M.~W. Schmeeckle, and A.~Badoux (2018), Resistance is not futile:
  Grain resistance controls on observed critical shields stress variations,
  \textit{Journal of Geophysical Research: Earth Surface}, \textit{123}(12),
  3308--3322, \doi{10.1029/2018JF004817}.

\bibitem[{\textit{Zhang et~al.}(2014)\textit{Zhang, Shao, and
  Huang}}]{Zhangetal14}
Zhang, J., Y.~Shao, and N.~Huang (2014), Measurements of dust deposition
  velocity in a wind-tunnel experiment, \textit{Atmospheric Chemistry and
  Physics}, \textit{14}, 8869--8882, \doi{10.5194/acp-14-8869-2014}.

\end{thebibliography}

\end{document}